# Pouch cells with 15% silicon calendar-aged for 4 years

Marco-Tulio F. Rodrigues[1]*, Zhenzhen Yang[1], Stephen E. Trask[1], Alison R. Dunlop[1], Minkyu Kim[1,2], Fulya Dogan[1], Baris Key[1], Ira Bloom[1], Daniel P. Abraham[1], Andrew N. Jansen[1]

[1] Chemical Sciences and Engineering Division, Argonne National Laboratory, Lemont, IL, USA

2 Current address: Department of Chemistry, Inha University, Incheon, 22212, Republic of Korea

***Corresponding author:** Marco-Tulio F. Rodrigues, marco@anl.gov

**Abstract**

Small amounts of high-capacity silicon-based materials are already used in the anode of commercial Li-ion batteries, helping increase their energy density. Despite their remarkable storage capability, silicon continuously reacts with the electrolyte, accelerating time-dependent cell performance fade. Nevertheless, very limited information is available on the specific consequences of this reactivity for the calendar aging of Li-ion cells. Here, we analyze aging effects on 450 mAh pouch cells containing 15 wt% of Si (and 73 wt% graphite) after storage at 21 °C for four years. We show that severe losses of Si capacity occurred due to particle isolation when cells were stored at high states of charge (SOC), but not when cells were fully discharged prior to storage. Impedance rise was also significantly higher when cells were kept at high SOCs and was mostly due to phenomena taking place at the cathode; the continuous electrolyte reduction at the anode did not lead to a major increase in bulk electrode resistance. A series of post-test characterization provided additional information on the effects of time and SOC on the calendar aging of Si-containing cells. Our study highlights the many challenges posed by Si during calendar aging and can inform future studies in the field.

*Keywords:* silicon, calendar aging, SEI growth, impedance rise, Si-graphite

**1. Introduction**

Silicon and Si-based materials have been widely explored as anode materials in Li-ion batteries, as their high capacity for Li$^+$ storage could improve the capabilities of these energy storage devices.[1] Lithiation of Si leads to the formation of a series of Li$_x$Si phases and significant restructuring of the Si-Si environments,[2] causing sizable changes in material density.[3] These density fluctuations manifest as large dimensional changes as Si-containing particles lithiate, initiating fractures that can accelerate performance fade.[1, 4] Years of research have focused on mitigating the consequences of this cyclic dilation/contraction, leading to < 5 wt% of Si-containing materials being deployed in some commercially available cells.[5, 6]

Despite this success, certain challenges remain unaddressed. Si-based materials and their lithiated forms display more *chemical* reactivity than graphite at equivalent electric potentials, causing decomposition of common polymeric binders and electrolyte solvents.[7-9] If left untamed, such reactivity can slowly deplete the electrolyte content in the cell and directly contribute to capacity fade. These mechanisms cause increased levels of performance decay due to the passage of time, even if cells are not actively cycled and the detrimental dimensional changes of Si particles are avoided. Indeed, a recent report indicated that insufficient calendar life in high-energy Si cells remains the biggest roadblock to their commercial deployment.[10]

Despite this clear challenge, there are few examples in open literature where time-dependent aging effects were investigated in cells containing Si-based materials. Zilberman et al.

reported a comprehensive study where 3.5 Ah cylindrical cells containing 3.5 wt% Si were stored for 11 months at 25 °C and various states-of-charge (SOCs); the team observed that, at low SOCs, capacity loss was dominated by parasitic reactions at the solid electrolyte interphase (SEI), but losses of Si capacity became relevant only above 30% SOC.[5] De Sutter et al. explored pouch cells (1.36 Ah) containing 55% of a Si alloy, which were stored at 100% SOC at both 25 °C and 45 °C; the cell at lower temperatures lost >10% of its capacity within 11 weeks of storage, while the one at 45 °C experienced severe capacity decay and gassing.[11] Zülke et al. investigated 4.8 Ah cylindrical cells containing 4.6 wt% of $SiO_x$, stored for one year at various temperatures and SOCs; in all cases, capacity fade was mostly dominated by reactions at the SEI, with little loss of $SiO_x$ capacity observed.[6] Sung et al. studied 110 Ah prismatic cells containing 5 wt% Si stored at 25% SOC for one year, in which cells retained 97.6% of their capacity; exact mechanisms of loss were not reported.[12] Finally, Lu et al. investigated coin cells containing 70 wt% $SiO_x$ tested intermittently for 80 days, and reported that capacity fade was mostly a function of time rather than cycling history.[13] Importantly, some of the works above report the observation of losses of accessible Si capacity due to calendar aging, which is typically not an active aging mechanism observed in Si-free graphite anodes.[5] Clearly, silicon behaves differently, and understanding how calendar aging is affected by the presence of silicon is of utmost importance for accurately forecasting the life of these new battery formulations.

The above reports on the time-dependent aging of Si-based cells generally focus on systems with low Si content that were evaluated for relatively short times with respect to the expected life of Li-ion batteries. Here, we describe the calendar aging behavior of 450 mAh pouch cells containing 15 wt% of Si nanoparticles during storage for *four years* at ambient temperature (21 °C). We demonstrate that significant losses of Si capacity can occur when cells are stored at high

SOCs, but not when cells are fully discharged prior to storage; further, we use various analytical tools to identify how calendar aging can affect the anode SEI and electrolyte composition. To the best of our knowledge, the analysis of cells with this high Si content and after such long aging is quite rare and hence can inform future developments in this field.

*2. Experimental*

*2.1. Materials and Methods.* The electrodes and cells used in this study were fabricated at Argonne's Cell Analysis, Modeling and Prototyping (CAMP) Facility. The positive electrode contained 90 wt% $LiNi_{0.6}Mn_{0.2}Co_{0.2}O_2$ (NMC622, ECOPRO), 5 wt% C45 conductive carbon (Timcal) and 5 wt% poly(vinylidene difluoride) binder (Solvay 5130). The negative electrode comprised 73 wt% MagE graphite (Hitachi), 15 wt% silicon nanoparticles (Nanostructured & Amorphous Materials, 50-70 nm), 2 wt% C45 conductive carbon and 10 wt% lithiated polyacrylic acid (LiPAA) binder. LiPAA was obtained from partial neutralization of PAA (Sigma-Aldrich, Mv ~450,000) with LiOH. The cathode and anode were coated on both faces of the aluminum and copper current collectors, respectively. After calendering, the thickness and estimated porosity for each coating side were 77 µm (34.7% porosity) and 53 µm (42.6%) for the cathode and anode, respectively. The areal loadings for each side of the electrode coating were ~3.6 mAh/cm$^2$ for the positive and ~4.5 mAh/cm$^2$ for the negative electrodes; the high N:P ratio limited anode lithiation to ~100 mV vs. Li/Li$^+$, decreasing the extent of Si volume expansion.

Pouch cells in the xx3450 format were built using twelve layers of electrodes, with 6 and 5 double-sided anode and cathode sheets, respectively; single-sided cathode layers were used at the extremes. The geometrical area of each layer was 14.1 cm$^2$ for the cathode and 14.9 cm$^2$ for the anode, and initial cell capacity was ~450 mAh. Celgard 2325 was used as separator, and the electrolyte was a 1.2 mol/L solution of LiPF$_6$ in a 3:7 wt:wt mixture of ethylene carbonate (EC) and ethylmethyl carbonate (EMC) (Tomiyama), with 10 wt% of fluoroethylene carbonate (FEC, Solvay). The electrolyte volume added to each cell amounted to 2.2x the total pore volume of electrodes and separator. Pouch cells were tested under a stack pressure of 15 kPa.

Cells were taken apart after testing and the electrolyte was collected for chemical analyses (see below). Electrochemical testing with harvested electrodes was performed using samples punched from layers in the center of the cell stack, with one side of the coating being gently removed using N-Methyl-2-pyrrolidone (NMP). Electrochemical post-testing was performed in 2032-format coin cells. Electrode capacity loss was estimated by punching 8 mm discs off the aged electrodes and assembling half-cells vs. Li; Celgard 2320 was used as separator. Electrochemical impedance spectroscopy (EIS) was carried out using symmetric cells containing two 8 mm discs of the same cycled electrode, spaced by a glass fiber separator; glass fiber was used in these cells due to its improved wettability by the electrolyte if compared to polyolefin separators. For these latter tests, electrodes were first individually equilibrated at 0.35 (anode) and 4.09 (cathode) V vs. Li/Li$^+$. The electrolyte composition was the same as that used in the pouch cells.

*2.2. Electrochemical testing.* Nine pouch cells were fabricated, with eight of them being promptly filled with electrolyte and the remainder being stored "dry"; this latter cell was used as a baseline to decouple the effects of calendar and cycle aging. After electrolyte filling, cells were exposed to a tap charge to 1.5 V, followed by 24 hours of rest. Formation cycles consisted of two cycles at C/20 and five cycles at C/10 (with a hold at 4.1 V until current decreased to C/16). After this initial conditioning, cells were degassed and resealed prior to additional testing. All electrochemical testing was performed at 30 ºC using 3.0 and 4.1 V as voltage cutoffs.

The conditioned cells were split into two sets, each aged using a different protocol. The *life protocol* (Figure 1a) consisted of three cycles at C/20, pulse-based impedance measurements (using 10-s discharge and charge pulses at 3C and 2.25C, respectively, collected at various voltages) and 92 cycles at C/3, followed by pulse impedance measurements and three final cycles at C/20. Cells were stored upon completion of this protocol, after being discharged to 3 V at C/20. A separate set of cells were tested using the *intermittent protocol* (Figure 1a), consisting of one cycle at C/20, pulse impedance measurement, two cycles at mixed rates (C/3 charge and C/2 discharge), and a C/20 charge to the aging voltage (initially 3.9 V, and later 4.0 V, as noted below). This final charge was followed by a 3-h voltage hold, after which cells were allowed to rest for 168 h (7 days). The protocol was repeated nine times (4x with aging at 3.9 V and 5x at 4.0 V), leading to a total test time of over 100 days. Cells that experienced this protocol were stored after being charged to 4.0 V (Figure 1a).

After ~4 years of storage in a dry room at ambient temperature (21 °C), cells were discharged at C/20 and then exposed to three full cycles at this same rate, followed by pulse impedance measurements and two cycles at mixed rate (see above). Cells were then charged to 3.9 V for EIS measurements. Cells were then discharged to 3 V for final storage and/or teardown; no electrode delamination was observed upon disassembly. EIS was carried out using a Solartron Analytical 1470E, with 5 mV amplitude and probing frequencies from 100 kHz to 10 mHz.

The pristine dry cell, previously under storage, was filled with electrolyte ~4.5 years after the completion of the initial *life* and *intermittent* experiments and was tested using the life protocol (Figure 1a). As shown below, performance was identical to that exhibited by cells filled immediately after assembly, validating the use of the dry cell as a baseline without calendar aging effects; storage of the dry cell without electrolyte did not cause visible aging.

Half-cells containing harvested electrodes were initially delithiated (for the anode) or relithiated (for the cathode), then cycled 3x at C/10 and once at C/100. Voltage cutoffs were 0.01 – 1.5 V and 3.0 – 4.3 V for these harvested anode and cathode cells, respectively. Experiments with harvested and symmetric cells were performed using discs punched from various regions of the aged electrodes, with most tests involving at least three cells. For clarity, only representative data is shown here.

*2.3. Physicochemical characterization.* Scanning electron microscopy (SEM) and energy-dispersive X-ray spectroscopy (EDS) were used to probe changes in electrode morphology and of compositional changes along the electrode matrix. The surface composition of harvested anodes was investigated using X-ray photoelectron spectroscopy (XPS). Electrolyte extracted from the aged cells was analyzed, and the identity of soluble, non-volatile electrolyte decomposition products was revealed using high-performance liquid chromatography (HPLC) coupled with electrospray-ionization MS (ESI/MS). Detailed information about instrumentation, sample preparation, experimental conditions and data analysis is provided elsewhere.[14-16]

Electrolyte harvested from cells was also analyzed using nuclear magnetic resonance (NMR) spectroscopy. Electrolyte extraction was achieved by soaking the pouch material and the separator in dimethyl carbonate; while this process ensures that electrolyte can be recovered even from highly aged cells, it also results in dilution, making it difficult to detect compounds present at low concentration. All sample preparation was carried out inside an argon-filled glovebox. A Bruker Avance III HD 300 MHz spectrometer was used to collect the $^{19}$F, $^{31}$P, $^{13}$C and $^{1}$H NMR spectra. The spectra were internally referenced to the signal of $PF_6^-$ at −72.7 ppm for $^{19}$F. For $^{31}$P, a 1M $H_3PO_4$ solution at 0 ppm was used. For $^{13}$C and $^{1}$H NMR, the spectra were referenced to TMS at 0 ppm. The electrolyte extracts were never exposed to air.

## 3. Results and Discussion

*3.1. Initial cell testing.* Cells tested under the *life* protocol lost ~40% of their initial discharge capacity at the end of 100 cycles (Figure 2a, shown in blue). For cells tested using the *intermittent* protocol, losses were linear and slightly faster on a per-cycle basis, likely due to the much longer time that elapsed during the experiment (Figure 2b, shown in red). Due to the different paths of aging, the final impedance was slightly higher for *intermittent* cells (Figure 2c); the larger impedance values at low voltages in aged electrodes is characteristic of Si.[17] After this initial testing, cells were stored at 21 °C for ~4 years, under conditions summarized in Figure 1a. Cells tested under the *life* protocol were stored for 4 years at 3 V, while cells tested under the *intermittent* protocol were stored at 4 V. Based on open circuit potentials measured in half-cells containing harvested electrodes and on the dQ/dV profile of full-cells, we estimate that the anode potentials during storage were ~650 mV and ~120 mV vs. Li/Li$^+$ for *life* and *intermittent* cells, respectively. Note that, by the time cells were first put in storage, *life* cells had a state of health (SOH, calculated vs. the initial cell capacity) of 60%, whereas *intermittent* cells had a SOH of 75%.

The pristine dry cell that was belatedly filled with electrolyte and tested under the *life* protocol exhibited performance that was indistinguishable from that exhibited by the initial batch (Figure 2, shown in orange). Hence, this cell provides a baseline for assessing the aging that is caused by cycling in contrast to that caused by cycling *and* time.

*3.2. Testing after 4 years of storage.* After this prolonged period of inactivity, cells underwent a reference performance test (RPT) including cycles at C/20 and impedance measurements. Representative discharge profiles measured at the beginning of testing (solid lines), before storage (dashed lines) and after 4 years of storage (dotted lines) are shown in Figure 3a for the *life* cells (stored at 3 V) and Figure 3b for the *intermittent* cells (stored at 4 V). Clearly, the aging paths for the two types of cells were markedly distinct: cells that were fully discharged prior to storage exhibited marginal capacity loss due to passage of time, whereas cells maintained at high voltage had their SOH decrease by an additional 13.3%. Furthermore, only a small variation in cell impedance was observed during storage in the discharged state (Figure 3c), while the impedance increased by ~50% when cells were stored at 4 V (Figure 3d). The extent of $Li^+$ inventory loss due to parasitic reactions at the SEI generally increases at lower anode potentials (higher cell SOCs).[18, 19] This is one likely reason for the higher capacity fade observed when cells were stored at higher voltages. The contribution of isolation of Si particles to these losses, and the sources of the impedance rise, are discussed in more detail below.

*3.3. Sources of impedance rise.* To better gauge the effects of calendar aging on cell impedance, EIS experiments were performed after the 4 years of storage (Figure 4a); data was collected after the RPT, with cells charged to 3.9 V. The "dry cell" was examined both before and after the aging cycles shown in Figure 2a and serves as a baseline for the initial characteristics of the cells.

After undergoing the *life* protocol, the spectrum for the dry cell exhibited an increase in the diameter for both semicircles (dashed orange line), indicating a deterioration in reaction kinetics after cycling. The cells previously cycled under this same protocol, and later stored at 3 V, underwent a mild additional increase in semicircle diameter (blue lines), suggesting that storage at low SOCs has limited effect on cell impedance. A very different scenario was observed for the *intermittent* cells (stored at 4 V), which exhibited sizable impedance rise, especially in the mid-frequency semicircle.

To quantify individual contributions to cell impedance rise, electrodes were punched from the cycled cells and assembled into symmetric cells; the impedance of a symmetric cell is twice that of the individual electrodes, providing a convenient method to determine individual electrode behavior.[20] Nyquist plots for the Si-Gr and the NMC622 electrodes are shown in Figures 4b-c, respectively. Spectra for symmetric cells using unaged electrodes are included in gray. The spectra for the anodes exhibit a broad semicircle, that displayed little change for cells stored at 3 V (blue). Anode impedance rise was more salient in the cells stored at 4 V (red), with certain cells exhibiting ~1.5x the original semicircle width of the unaged electrodes. Regardless of cell history, the low-frequency tail of all calendar-aged cells extended to higher impedance values, suggesting that $Li^+$ diffusion within the anode has deteriorated. Nevertheless, the absolute changes in anode impedance were significantly smaller than the observed for the cathode (Figure 4c), which doubled after the *life* protocol and storage at 3 V (blue) and almost quadrupled after storage at high SOC (red). Most changes in cathode impedance were observed in the mid-frequency range, in agreement to what is observed in the full-cells (Figure 4a). Hence, despite the expected role of parasitic reactions at the SEI on the capacity fade during calendar aging (Figure 3), most changes to cell impedance originated at the cathode. This prominent role of the positive electrode on cell impedance rise is not uncommon, and has been associated with oxygen loss from the particle surface and with the deposition of oxidized matter around particles.[21] As these processes are promoted by high voltages, they are more prominent after long-term storage at high SOCs (Figure 4c).

Another interesting feature visible in Figure 4a is the increase in the high-frequency intercept as function of aging conditions. After formation, the intercept occurs at 6.6 $\Omega$ $cm^2$, and it increases to 7.8 $\Omega$ $cm^2$ after the *life* protocol (dashed orange line). After four years, intercepts were observed at ~9.8 Ω cm$^2$ for storage at 3 V, and ~14.9 Ω cm$^2$ for storage at 4 V. All else constant, changes to this intercept could be indicative of an increase in electrolyte resistance within the cell stack,[22] such as due to the formation of gas pockets after electrolyte decomposition. The interpretation that this feature arises from the electrolyte is supported by the spectra in Figures 4b-c, in which changes in the intercept are absent. As expected, this increase in electrolyte resistance is significantly more problematic for cells stored a higher SOCs, as the driving forces for both oxidative and reductive parasitic processes at cathode and anode, respectively, become larger. In fact, more gassing was observed in the *intermittent* than in *life* cells (Figures 1b-c), in agreement with a larger extent of electrolyte decomposition. Electrolyte depletion can inactivate entire portions of the electrode due to de-wetting,[23] and can accelerate performance fade. We note that the cells used in this study contained an electrolyte volume equal to 2.2x the total pore volume of electrodes and separator, which is likely in excess of what is found in commercial cells. Thus, it is possible that our observations underestimate the effect of calendar aging on the cells, as Figure 4a indicates that electrolyte depletion could be a relevant aging mode in these cells. Accelerated electrolyte reactivity with silicon has been suggested to be problematic for the longevity of cells with high Si content.[10]

*3.4. Quantifying active material capacity loss.* A previous report analyzing the calendar aging of cells containing small amounts of silicon in the anode (< 5%) concluded that permanent losses of Si capacity were the main source of capacity fade at most SOCs.[5] To assess the losses of silicon in our cells, we extracted harvested electrodes and tested them in half-cells (vs. Li metal). Half-cells with unaged (pristine) electrodes were also tested for reference. Lithiation and delithiation profiles measured at C/100 are shown in Figures 5a-b, respectively. After aging, a ~17% loss of anode capacity was observed, regardless of testing and storage conditions. Inspection of delithiation profiles (Figure 5b) suggest that this loss can be *entirely* attributed to silicon. Due to the large voltage hysteresis of silicon, Li$^+$ extraction from graphite progresses almost to its full extent before Si begins to be delithiated.[24, 25] The plateaus that are typical of graphite intercalation show great overlap between aged and unaged samples in Figure 5b, demonstrating that *the electrochemical activity of graphite particles remained undisturbed* even after long-term storage of the cell. Conversely, the portions of the voltage profile bearing solely the response of silicon are visibly shortened, indicating a decrease in the number of Si particles that remain electrochemically active in the electrode.

We can use the information above to estimate how much of the initial capacity of silicon is lost during the test. Assuming that all the delithiation capacity below 250 mV vs. Li/Li$^+$ is related to graphite and the remainder to silicon, the unaged electrode presented ~1.74 mAh/cm$^2$ of graphite capacity and ~2.9 mAh/cm$^2$ of Si (Figure 5b). After aging, this latter component decreases to ~1.94 mAh/cm2. That is, while aging caused the overall electrode capacity to decrease (on average) by 17%, about 33% of the silicon particles in the anode are no longer accessible to Li$^+$ ions.

Interestingly, Figure 5b also shows a drastic change in the phase behavior of silicon in the anode. The unaged electrodes exhibited a well-defined delithiation plateau at ~400 mV vs. Li/Li$^+$, associated with lithium extraction from a crystalline Li$_{15}$Si$_4$ phase. [1, 26] In all aged samples, however, this plateau was nearly completely absent, and the silicon domains behaved as amorphous Li$_x$Si compounds. This shift also caused subtle changes to the delithiation profile at ~200 mV vs. Li/Li$^+$, where the unaged electrodes presented a steeper transition in between the two plateaus of graphite. Amorphous Li$_x$Si can be delithiated at lower overpotentials than the crystalline Li$_{15}$Si$_4$ and contributes to the delithiation capacity at that potential range. Since the coulombic efficiencies for the C/100 cycle were similar in all half-cells, it is unlikely that crystalline domains become inactive after being formed in the aged samples. Rather, we believe that the crystalline phase is unable to be fully nucleated in the aged electrodes. Formation of the $Li_{15}Si_4$ phase is usually observed at ~50 mV vs. $Li/Li^+$, but differences in the strain state of Si particles caused by changes in size and/or morphology can increase the overpotential needed for this transformation, preventing it from happening altogether under certain voltage limits.[27] During lithiation (Figure 5a), a small plateau at this potential can be seen in all samples. However, unaged electrodes also present a smaller plateau at 30 mV vs. $Li/Li^+$, which is absent from the harvested electrodes. The absence of this latter transition could be the reason for the lack of crystallinity after lithiation of silicon particles in the aged electrodes. We hypothesize that excessive fragmentation of Si due to cyclic dimensional changes leads to effective particle sizes that are much smaller than at the beginning of the test, affecting the crystallization overpotentials.

In summary, aging of the cells caused ~33% of Si particles to become unresponsive, while also changing the phase behavior as the remaining domains are lithiated. Such evolution in the qualitative features of the profiles could pose a challenge to some battery control algorithms. Remarkably, graphite domains remain unscathed, retaining their activity despite the significant inactivation of Si particles. Different from the anode, cells containing discs of NMC622 cathode harvested from the aged cells did not show significant capacity loss (Figure 5c).

*3.5. Differentiating cycling losses from storage losses.* Figures 5a-b show that losses of silicon capacity were similar for cells stored at either 3 V or 4 V for a total of four years. However, recall that each set of cells also experienced a different testing protocol prior to storage (Figure 1a), and that they entered the storage period at different states-of-health (Figure 3a). Hence, the information provided by harvested electrodes is incomplete, as it does not convey how much of the losses occurred during initial testing, and how much was solely due to calendar aging while in storage.

Since the dry cell was exposed to the *life* protocol and immediately disassembled for post testing, it can be used as a baseline to gauge calendar aging effects on the cells stored at 3 V. Inspection of Figures 5a-b shows that losses observed in *life* and dry cells were in fact very similar. That is, a cell that was simply exposed to one hundred cycles lost about as much Si capacity as cells exposed to this protocol *and* stored at 3 V and 21 °C for 4 years. Thus, *storage at low states-of-charge appears to only cause marginal losses of Si capacity*. This is in agreement with the work by Zilberman et al.,[5] which found that losses of $Li^+$ inventory to the SEI in cells with 3.5 wt% Si remained the dominant source of capacity fade after 11 months at low SOCs.

Although we do not have a similar baseline for the *intermittent* cells that would allow us to directly *measure* how much of the Si loss was caused due to storage, we can achieve that through analysis of dQ/dV profiles. The characteristic plateaus related to the staging behavior of graphite (Figures 5a-b) give rise to sharp peaks in the dQ/dV profile. Some of these graphite peaks are often conspicuous when analyzing dQ/dV profiles of *full-cells*, enabling the identification of periods of graphite activity in blended electrodes containing silicon. This, combined with the knowledge of the range of potentials in which Si lithiates (Figure 5a), allow us to infer losses of Si capacity from the full-cell data.

Figure 6a shows the dQ/dV profile during charge of pouch cells tested using the cycling protocol and stored at 3 V, plotted as a function of cell capacity instead of the more typical representation vs. voltage. The solid line indicates the data at the beginning of life, the dashed line that prior to storage, and the dotted profiles the behavior after 4 years. The prominent peak indicated by the • arises from the two-phase region of graphite characterized by the plateau at ~210 mV vs. $Li/Li^+$. Recalling the discussion in Figures 5a-b, silicon particles begin lithiating at higher potentials than graphite, and thus all capacity prior to the peak is related to silicon (though not all silicon capacity is accessed at this point, as particles continue to lithiate when graphite becomes electrochemically active). The gray vertical lines in Figure 6 are a guide to the eye for the amount of capacity that is contributed by silicon before graphite begins lithiating. Immediately after formation, cells tested under the *life* protocol accept 0.95 mAh/cm$^2$ before the onset of graphite lithiation. Right before storage, this region of early Si activity shrinks to 0.73 mAh/cm$^2$. Little change is observed after four years of storage at 3 V, suggesting that the *inactivation of Si particles occurred primarily due to cycling, but not due to long-term storage at low SOCs*. This assessment agrees remarkably well with the data in Figures 5a-b, in which electrodes harvested from the *life* and dry cells exhibited similar losses.

The aging of the *intermittent* cells tells a very different story (Figure 6b). Here, graphite lithiation initially commences after 0.93 mAh/cm$^2$, changing only slightly to 0.88 mAh/cm$^2$ when measured right before storage. However, after four years of storage at 4 V, this number decreases to 0.75 mAh/cm$^2$. That is, for these cells, *much larger losses of Si capacity occurred due to storage at high SOCs* than due to preliminary testing. Although the final state of Si losses in the two sets of cells is similar, storage at high SOCs is much more aggressive and consequential than at low SOCs.

This SOC-dependence on the losses of Si capacity during calendar aging indicates that the underlying processes are electrochemical in nature. It has been shown that the extensive growth of SEI could cause isolation of silicon particles.[28] Since the driving force for reduction reactions increases at low anode potentials (i.e., at high cell SOCs),[18, 19] the higher rate of side reactions that also caused more capacity fade during storage (Figure 3a) also contributed to larger losses of active material. Note that these losses are due to the intrinsic reactivity of Si particles instead of fracturing, as little dimensional changes are expected during storage. Storage SOC has been shown to be important to minimize loss of Li$^+$ inventory in graphite,[18, 19] but in Si the consequences are much more drastic, as active material capacity is also lost in the process.

Interestingly, we can also use this SOC-dependency to rule out the contribution of certain mechanisms to this loss of active material capacity. The surface of silicon particles is susceptible to attack by hydrofluoric acid,[29, 30] which is formed from reactions with labile fluorine from the LiPF$_6$ salt.[31] An extreme example in the literature shows that a Si wafer can be directly etched by this HF.[32] Since the acidolysis of Si is a chemical process, its rate would depend more on the concentration of HF in the electrolyte rather than cell (and anode) potential. Given the long duration of static storage of the cells, the negligible capacity loss observed at low SOC suggests that *direct etching by HF is unlikely to be a relevant source of Si capacity loss during extended storage*.

As a final note on the method illustrated in Figure 6, although estimates of Si loss show good quantitative agreement with values extracted from studies with harvested electrodes (Figures 5a-b), it does not provide a perfect match. Various aging modes (including Li$^+$ inventory loss) cause changes in the end-of-discharge potential of the anode,[33] leading to small alterations of how much capacity silicon can contribute before graphite starts lithiating. Additionally, we note that applying this method to discharge (instead of charge) can be complicated due to changes in the final state of lithiation of silicon as the cell ages.

*3.6. Self-discharge during storage.* With the knowledge of how electrodes have calendar-aged in each type of cell, we now turn our attention to the self-discharge behavior during the four years of storage. Figure 7a shows a hysteresis plot of the voltage profiles of the cell stored at 4 V, showing the final charge prior to storage and the initial discharge four years later. After four years, the voltage decreased to 3.81 V as the cells lost ~21% of capacity. This voltage drop during storage is typically caused by a combination of oxidative and reductive parasitic processes. While the latter causes losses of $Li^+$ inventory, oxidation at the cathode actually injects extra $Li^+$ inventory into the cell,[21, 34] leaving a lower number of empty sites in the cathode to be filled during the ensuing discharge. For that reason, it is not unusual for this initial discharge to overestimate the $Li^+$ consumption after extended storage, and additional cycles are recommended for a more accurate quantification. Interestingly, for our cells, the capacity loss measured from this first cycle was actually smaller than that measured from additional full cycles (with 3 V and 4.1 V as cutoffs). We attribute this behavior to the losses of Si capacity discussed in Figure 6.

Figure 7b shows the dQ/dV profiles (vs. cell capacity) of the charge prior to the storage and the subsequent discharge obtained four years later. During charge, the peak related to the graphite plateau is visible halfway through, suggesting that the anode was lithiated to at least ~120 mV vs. $Li/Li^+$ prior to long-term storage. During discharge, only broad features are visible. It is not clear whether graphite remained lithiated at all, and the cell may likely have self-discharged to anode potentials >200 mV vs. $Li/Li^+$. Considering that capacity fade during calendar aging is inversely proportional to the anode potential, all effects we discuss here could be even more severe if the cells had been stored at even higher SOCs.

Interestingly, the unusually steep onset of the discharge profile in Figure 7a was consistently observed for all four cells stored at 4 V. We have previously observed lithiated Si particles "freezing" during experiments, becoming almost completely unresponsive to delithiation unless some overpotential is overcome.[35] We hypothesize that this could be another manifestation of such behavior, causing an additional 60 mV drop in cell voltage.

Now we turn our attention to the *life* cells, discharged to 3 V prior to storage. One minute after discharging (when the final data point was collected), the OCV rebounded to 3.11 V. Four years later, it was 3.18 V. The limited amount of data acquired before storage makes it difficult to exactly assess how cell voltage has evolved over time. Nevertheless, cells clearly remained away from overdischarge even after four years of storage. Based on the analysis of harvested electrodes, anode potentials were estimated to be ~650 mV vs. Li/Li$^+$ when cells were torn down. Although it is unknown if Si-free cells would have gone into overdischarge after such long storage under similar conditions, we believe that the presence of Si could also have acted as a deterrent. Graphite has a steep voltage profile at very low SOCs, causing the anode potential to rise abruptly after minute charge loss. Silicon, on the other hand, has a much smoother profile, spanning a voltage range that is wider than the window typically cycled within full-cells; that is, 0% SOC in a full-cell occurs at a non-negligible SOC in the anode. This "excess" capacity could presumably help maintain anode potentials within safe levels after moderate capacity fade during extended periods. Overdischarge is reported to cause oxidation of the SEI and dissolution of the copper current collector, which are detrimental to cell health.[36, 37] Evaluation of this presumed protective effect of silicon could be the focus of future studies.

*3.7. Physical characterization of electrodes and electrolyte.* Irreversible changes in electrode thickness due to aging were evaluated using scanning electron microscopy. Figure S1a shows a cross-section image of the pristine Si-graphite electrode; the initial total thickness of the double-sided electrode was 110 µm. For the dry cell that was freshly exposed to the *life* protocol (no calendar aging, Figure S1b), the anode showed irreversible dilation to 170 µm. Note that the electrode was harvested from a fully discharged cell, and thus this thickness change disregards cyclic contributions arising from the dilation of the lithiated silicon particles. Cells tested with the *life* protocol and stored at 3 V for four years showed only a minor additional dilation compared to the dry cell (Figure S1c), in agreement with the low levels of capacity fade and Si isolation discussed above. Anodes harvested from cells tested under the intermittent protocol and stored at 4 V for four years showed a final thickness of 160 µm (Figure S1d). In this latter case, it is difficult to dissociate the effect of previous cycling history from that of aging. Nevertheless, the fact that the thickness remains smaller than that of cells that were exposed to more extensive cycling indicates that the *irreversible dilation is more pronounced as a consequence of cycling than of prolonged storage*.

Energy-dispersive X-ray spectroscopy was used to map the distribution of elements across the electrode cross-sections (Figure 8). In all cases, signals associated with the active materials (silicon and graphite) remained distributed across the entire thickness of the electrode, indicating that thickness changes occurred mostly due to uniform expansion of the entire electrode matrix rather than due to surface deposition of decomposed matter. Closer inspection of the signals related to SEI components (oxygen, fluorine and phosphorus) shows that they appear to be more correlated with the spatial distribution of silicon than graphite (especially for the dry cell), suggesting that most SEI products tend to accumulate around silicon. This observation agrees with the noted losses of silicon capacity but not of graphite, as the former becomes isolated by the electrolyte decomposition products.

Compositional analysis of the anode surface after cell teardown was carried out using XPS (Figure S2). In general, it appears that storage for four years led to the formation of a more mineralized layer, as evidenced by the enrichment of LiF in the SEI. There are also less fluorophosphates, which are formed during hydrolysis of the $LiPF_6$ salt, indicating that more advanced stages of hydrolysis ($Li_xPO_y$) could be encountered after long-term aging. Additionally, all these transformations were more pronounced in cells stored at 4 V, possibly because of a more reducing environment at the anode. Interestingly, $Li_xC_6$ (~282 eV) is detected at much higher intensities in the cell stored at 4 V, indicating incomplete delithiation of graphite domains when cells were discharged prior to disassembly. This apparent isolation of some graphite domains is unlikely to be permanent, as all graphite capacity were observed to remain accessible at C/100 (Figures 5a-b).

Analysis of the electrolyte extracted from the aged cells using HPLC/MS-ESI and NMR is discussed in the *Supplementary Material*. Just as observed using XPS, the cell SOC during storage did not have a strong influence on the final electrolyte composition.

*4. Conclusions*

The calendar aging behavior of Si-based Li-ion cells was investigated after 4 years of storage at 21 °C. The two sets of 450 mAh pouch cells containing 15 wt% of silicon nanoparticles in the anode differed both in their previous testing history and in their storage conditions. Cells tested under the *life* protocol underwent a total of 100 aging cycles, and were stored after a full discharge to 3 V; cells tested under the *intermittent* protocol experienced 40 aging cycles and were stored at 4 V. The main takeaways from this study are summarized below.

- Storing cells at 4 V (anode at ~120 mV vs. Li/Li$^+$) led to much higher levels of capacity fade due to calendar aging than when cells were stored at 3 V (anode at ~650 mV vs. Li/Li$^+$). Part of this difference comes from the higher driving force for electrolyte reduction at lower anode potentials. Additionally, severe losses of Si capacity were observed for the cells stored at high SOC but were negligible for the fully discharged cells.
- Permanent losses of Si capacity were a consequence of an electrochemical rather than a chemical process. Consequently, direct dissolution of silicon by HF is unlikely to be a relevant aging mechanism. Rather, it appears that Si domains become electrically isolated as some electrode pores are clogged by sustained electrolyte decomposition.

- The surface layer of the anode became more inorganic in character after extended storage, and even more so for cells maintained at high SOCs. Additionally, analysis of electrolyte recovered from the cells identified similar decomposition products, regardless of the SOC of storage.
- Impedance rise was much larger in cells stored at high SOCs and occurred primarily due to processes on the cathode. Despite the continuous $Li^+$ losses to the SEI, changes to the bulk impedance of the anode were comparatively small.
- The high frequency intercept of the impedance spectra shifted towards higher values after storage, especially for cells kept at high SOC. Such shifts are likely correlated with higher levels of electrolyte consumption, as evidenced by visible bulging of the pouch cells. Since the cells used in this study contained a slight electrolyte excess, the real effects of electrolyte depletion on aging are underestimated and are expected to be more severe in "electrolyte-limited" commercial cells.
- The irreversible increase in anode thickness was mostly caused by cycling rather than calendar aging.
- Formation of the crystalline $Li_{15}Si_4$ phase is kinetically frustrated in aged electrodes, causing drastic changes to the delithiation profile of Si-graphite anodes. Such changes could affect diagnostic algorithms reliant on the analysis of these profiles.

In addition to these observations, our analysis also suggested the presence of Si-based materials could potentially help protect cells from overdischarge during extended storage at very low SOCs, meriting further investigation. We believe that the observations and ideas contained in this work can help support the commercial deployment of Si-containing Li-ion batteries.

*Acknowledgements*

This research was supported by the U.S. Department of Energy's Vehicle Technologies Office under the Silicon Consortium Project, directed by Brian Cunningham, and managed by Anthony Burrell. The submitted manuscript has been created by UChicago Argonne, LLC, Operator of Argonne National Laboratory ("Argonne"). Argonne, a U.S. Department of Energy Office of Science laboratory, is operated under Contract No. DE-AC02-06CH11357. The U.S. Government retains for itself, and others acting on its behalf, a paid-up nonexclusive, irrevocable worldwide license in said article to reproduce, prepare derivative works, distribute copies to the public, and perform publicly and display publicly, by or on behalf of the Government.

*References*

[1] M.N. Obrovac, V.L. Chevrier, Chemical Reviews, 114 (2014) 11444-11502.

[2] B. Key, R. Bhattacharyya, M. Morcrette, V. Seznéc, J.-M. Tarascon, C.P. Grey, Journal of the American Chemical Society, 131 (2009) 9239-9249.

[3] B.A. Boukamp, G.C. Lesh, R.A. Huggins, Journal of The Electrochemical Society, 128 (1981) 725-729.

[4] X.H. Liu, L. Zhong, S. Huang, S.X. Mao, T. Zhu, J.Y. Huang, ACS Nano, 6 (2012) 1522-1531.

[5] I. Zilberman, J. Sturm, A. Jossen, Journal of Power Sources, 425 (2019) 217-226.

[6] A. Zülke, Y. Li, P. Keil, R. Burrell, S. Belaisch, M. Nagarathinam, M.P. Mercer, H.E. Hoster, Batteries & Supercaps, 4 (2021) 934-947.

[7] B. Han, Y. Zhang, C. Liao, S.E. Trask, X. Li, R. Uppuluri, J.T. Vaughey, B. Key, F. Dogan, ACS Applied Materials & Interfaces, 13 (2021) 28017-28026.

[8] B. Han, M.J. Piernas-Muñoz, F. Dogan, J. Kubal, S.E. Trask, I.D. Bloom, J.T. Vaughey, B. Key, Journal of The Electrochemical Society, 166 (2019) A2396-A2402.

[9] C.L. Seitzinger, R.L. Sacci, J.E. Coyle, C.A. Apblett, K.A. Hays, R.R. Armstrong, A.M. Rogers, B.L. Armstrong, T.H. Bennet, N.R. Neale, G.M. Veith, Chemistry of Materials, 32 (2020) 3199-3210.

[10] J.D. McBrayer, M.-T.F. Rodrigues, M.C. Schulze, D.P. Abraham, C.A. Apblett, I. Bloom, G.M. Carroll, A.M. Colclasure, C. Fang, K.L. Harrison, G. Liu, S.D. Minteer, N.R. Neale, G.M. Veith, C.S. Johnson, J.T. Vaughey, A.K. Burrell, B. Cunningham, Nature Energy, 6 (2021) 866-872.

[11] L. De Sutter, G. Berckmans, M. Marinaro, J. Smekens, Y. Firouz, M. Wohlfahrt-Mehrens, J. Van Mierlo, N. Omar, Energies, 11 (2018) 2948.

[12] J. Sung, N. Kim, J. Ma, J.H. Lee, S.H. Joo, T. Lee, S. Chae, M. Yoon, Y. Lee, J. Hwang, S.K. Kwak, J. Cho, Nature Energy, 6 (2021) 1164-1175.

[13] W. Lu, L. Zhang, Y. Qin, A. Jansen, Journal of The Electrochemical Society, 165 (2018) A2179-A2183.

[14] Z. Yang, J.W. Morrissette, Q. Meisner, S.-B. Son, S.E. Trask, Y. Tsai, S. Lopykinski, S. Naik, I. Bloom, Energy Technology, 9 (2021) 2000696.

[15] M.J. Piernas-Muñoz, Z. Yang, M. Kim, S.E. Trask, A.R. Dunlop, I. Bloom, Journal of Power Sources, 487 (2021) 229322.

[16] R. Sahore, F. Dogan, I.D. Bloom, Chemistry of Materials, 31 (2019) 2884-2891.

[17] M. Klett, J.A. Gilbert, S.E. Trask, B.J. Polzin, A.N. Jansen, D.W. Dees, D.P. Abraham, Journal of The Electrochemical Society, 163 (2016) A875-A887.

[18] P. Keil, S.F. Schuster, J. Wilhelm, J. Travi, A. Hauser, R.C. Karl, A. Jossen, Journal of The Electrochemical Society, 163 (2016) A1872-A1880.

[19] F. Single, A. Latz, B. Horstmann, ChemSusChem, 11 (2018) 1950-1955.

[20] C. Bünzli, H. Kaiser, P. Novák, Journal of The Electrochemical Society, 162 (2014) A218-A222.

[21] M.-T.F. Rodrigues, K. Kalaga, S.E. Trask, I.A. Shkrob, D.P. Abraham, Journal of The Electrochemical Society, 165 (2018) A1697-A1705.

[22] K. Mc Carthy, H. Gullapalli, K.M. Ryan, T. Kennedy, Journal of The Electrochemical Society, 168 (2021) 080517.

[23] Z. Deng, Z. Huang, Y. Shen, Y. Huang, H. Ding, A. Luscombe, M. Johnson, J.E. Harlow, R. Gauthier, J.R. Dahn, Joule, 4 (2020) 2017-2029.

[24] K.P.C. Yao, J.S. Okasinski, K. Kalaga, J.D. Almer, D.P. Abraham, Advanced Energy Materials, 9 (2019) 1803380.

[25] W. Ai, N. Kirkaldy, Y. Jiang, G. Offer, H. Wang, B. Wu, Journal of Power Sources, 527 (2022) 231142.

[26] M.N. Obrovac, L. Christensen, Electrochemical and Solid-State Letters, 7 (2004) A93.

[27] D.S.M. Iaboni, M.N. Obrovac, Journal of The Electrochemical Society, 163 (2015) A255-A261.

[28] Y. He, L. Jiang, T. Chen, Y. Xu, H. Jia, R. Yi, D. Xue, M. Song, A. Genc, C. Bouchet-Marquis, L. Pullan, T. Tessner, J. Yoo, X. Li, J.-G. Zhang, S. Zhang, C. Wang, Nature Nanotechnology, 16 (2021) 1113-1120.

[29] K. Kalaga, M.-T.F. Rodrigues, S.E. Trask, I.A. Shkrob, D.P. Abraham, Electrochimica Acta, 280 (2018) 221-228.

[30] J. Bareño, I.A. Shkrob, J.A. Gilbert, M. Klett, D.P. Abraham, The Journal of Physical Chemistry C, 121 (2017) 20640-20649.

[31] M. Stich, M. Göttlinger, M. Kurniawan, U. Schmidt, A. Bund, The Journal of Physical Chemistry C, 122 (2018) 8836-8842.

[32] Y. Ha, C. Stetson, S.P. Harvey, G. Teeter, B.J. Tremolet de Villers, C.-S. Jiang, M. Schnabel, P. Stradins, A. Burrell, S.-D. Han, ACS Applied Materials & Interfaces, 12 (2020) 49563-49573.

[33] M. Dubarry, C. Truchot, B.Y. Liaw, Journal of Power Sources, 219 (2012) 204-216.

[34] A. Tornheim, D.C. O'Hanlon, Journal of The Electrochemical Society, 167 (2020) 110520.

[35] M.-T.F. Rodrigues, J.A. Gilbert, K. Kalaga, D.P. Abraham, Journal of Physics: Energy, 2 (2020) 024002.

[36] D. Juarez-Robles, A.A. Vyas, C. Fear, J.A. Jeevarajan, P.P. Mukherjee, Journal of The Electrochemical Society, 167 (2020) 090558.

[37] K.R. Crompton, B.J. Landi, Energy & Environmental Science, 9 (2016) 2219-2239.

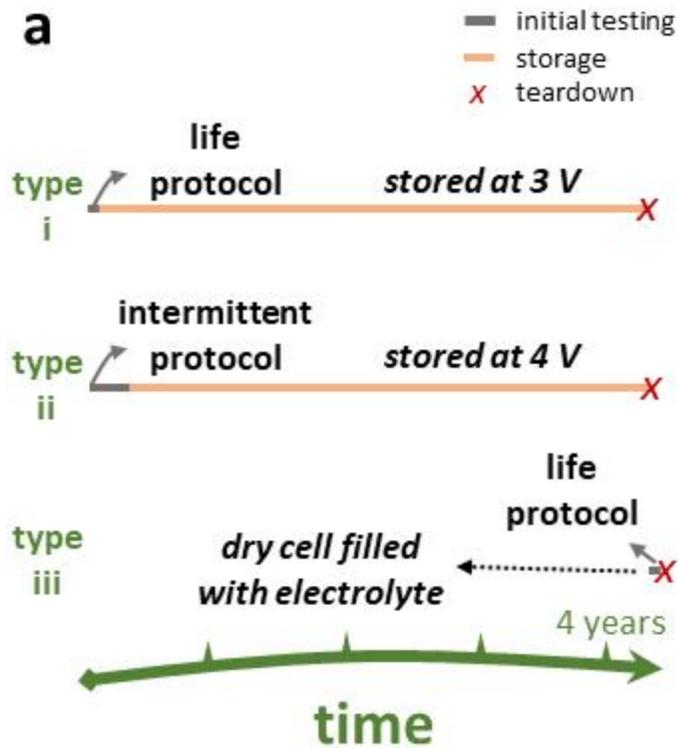
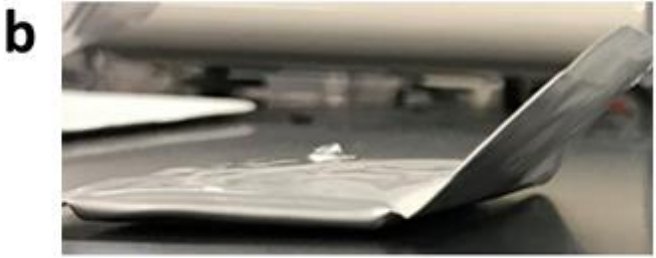
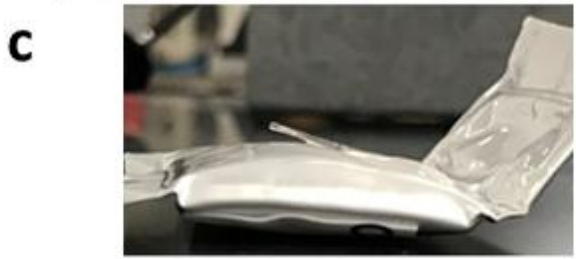

Figure 1. Overview of testing conditions. a) Schematic representation of initial cell testing and calendar aging, highlighting the final cell voltages prior to storage. b) Photograph of a *life* cell after 4 years of storage at 3 V. c) Photograph of an *intermittent* cell after 4 years at 4 V. More gassing was observed in cells stored at higher voltages.

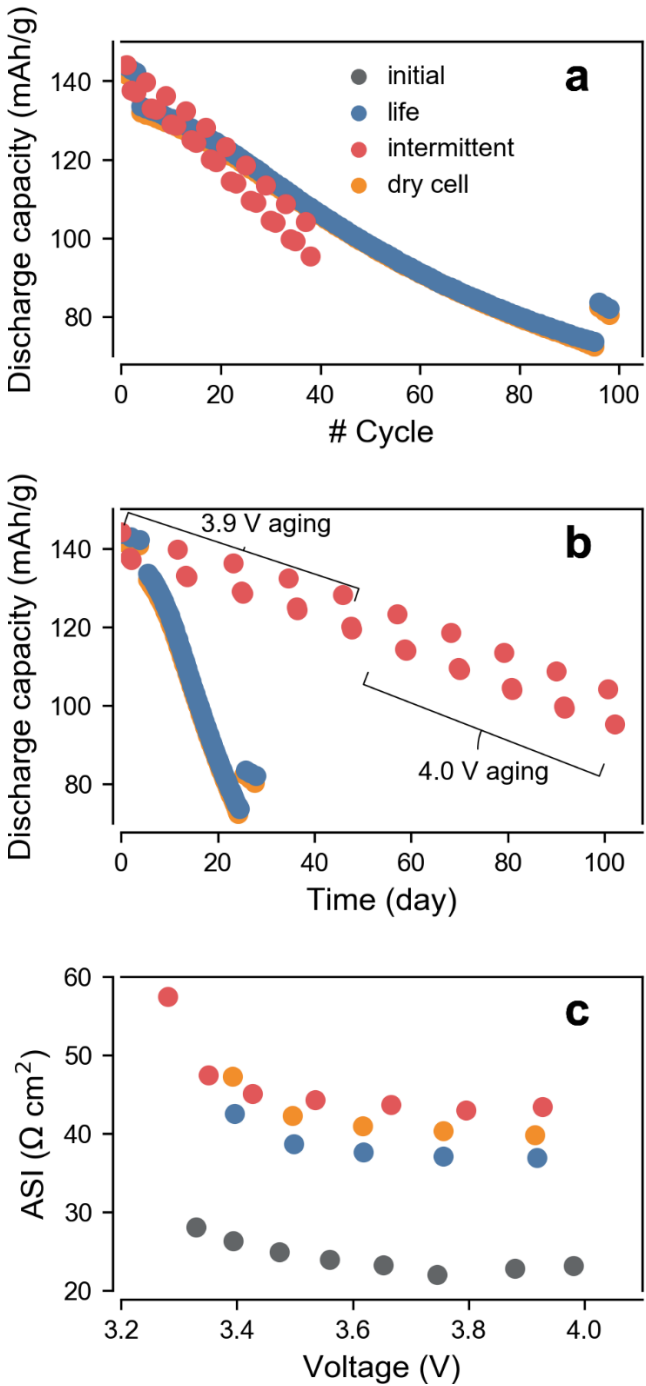

Figure 2. Cell testing prior to storage. a) Discharge capacities as a function of cycle number. b) Discharge capacities as a function of total test time. c) Area-specific impedances (ASI) measured using discharge pulses. Panel b indicates the voltages to which *intermittent* cells were charged

prior to each 1-week rest period; discharge capacities immediately after each rest are omitted for clarity. Symbols indicate average values, and the standard deviations were smaller than symbol size. The legend in panel a applies to all panels.

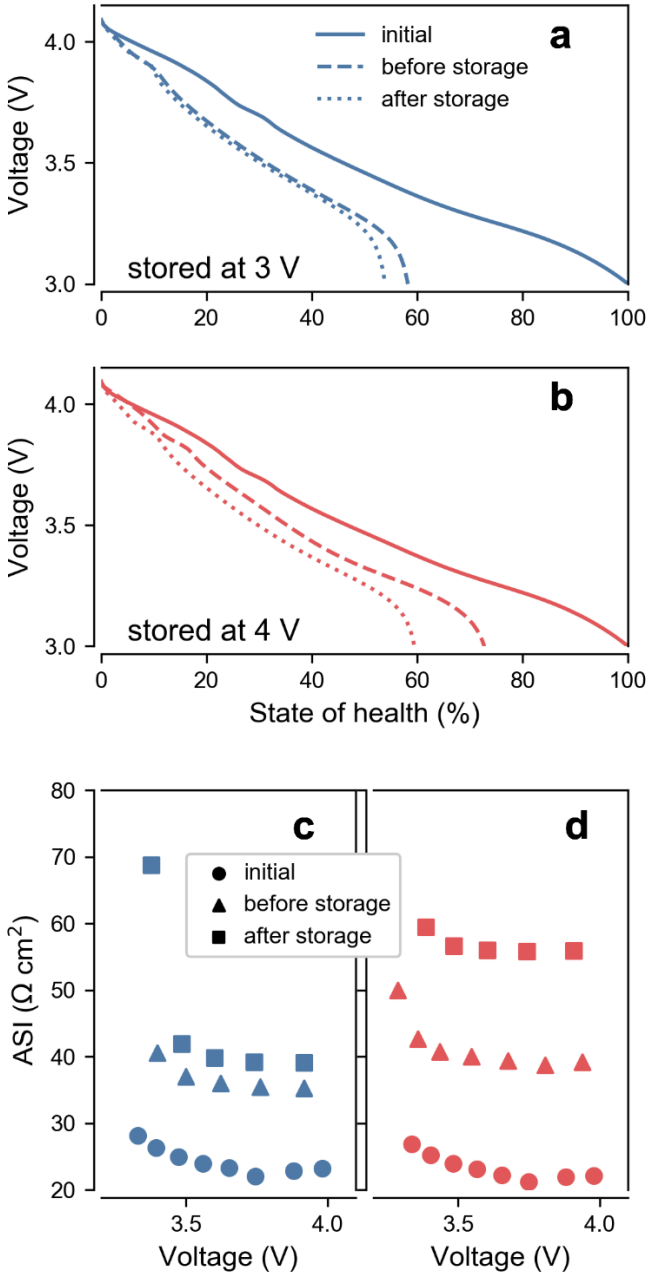

Figure 3. Cell testing after 4 years of storage. a) Representative discharge profiles for *life* cells at various points of testing. b) Representative discharge profiles for *intermittent* cells at various points of testing. Area-specific impedances measured at each of these points are shown in (c) for *life* and (d) for *intermittent* cells.

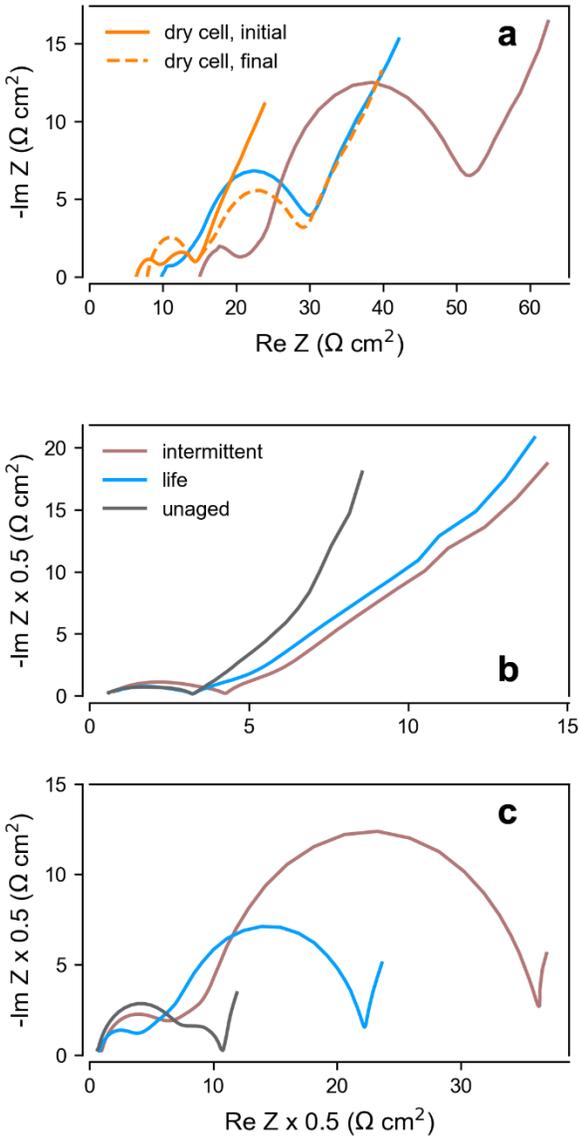

Figure 4. Representative electrochemical impedance spectra. a) Impedance spectra of pouch cells equilibrated at 3.9 V. b) Impedance spectra of anode vs. anode symmetric cells, with each electrode individually pre-lithiated to 0.35 V vs. Li metal. c) Impedance spectra of cathode vs. cathode symmetric cells, with each electrode individually pre-delithiated to 4.09 V vs. Li metal. In panels b and c, ASI values were halved to indicate the equivalent single-electrode quantities. The color code in panel b applies to all panels.

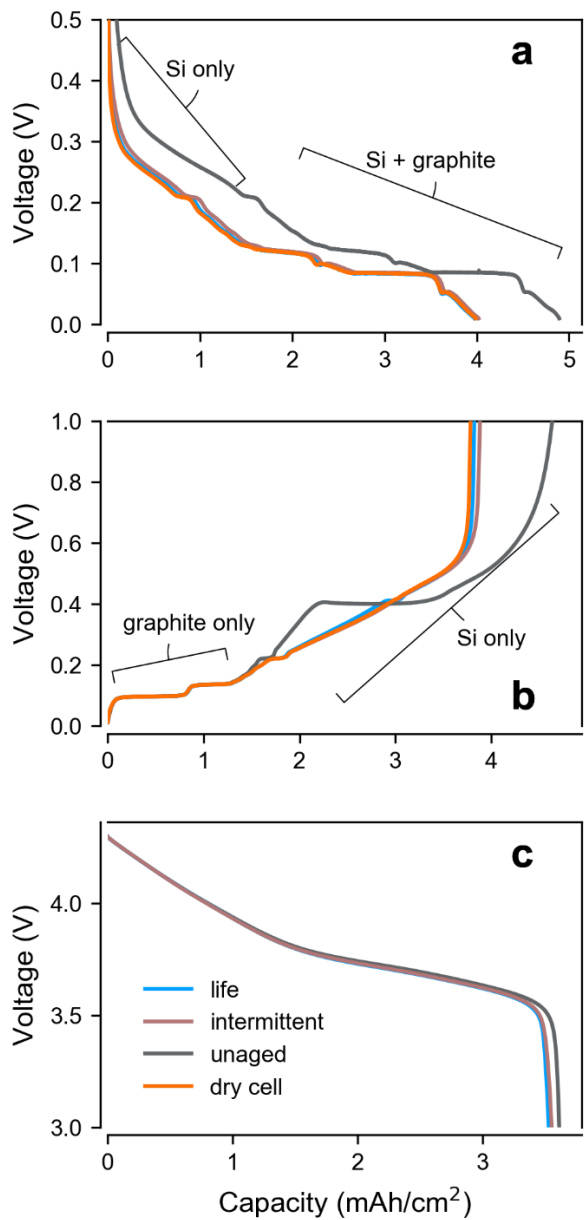

Figure 5. Representative voltage profiles measured at C/100 for half-cells using harvested electrodes. a) Lithiation profiles for Si-graphite anodes. b) Delithiation profiles for Si-graphite anodes. c) Relithiation profiles for NMC622 cathodes. Permanent losses are much more prominent in harvested anode than cathode samples, and little variation was observed in the final state of the electrodes with respect to their aging path. The brackets in panels a and b indicate the regions of activity of each anode component. The color code in panel c applies to all panels.

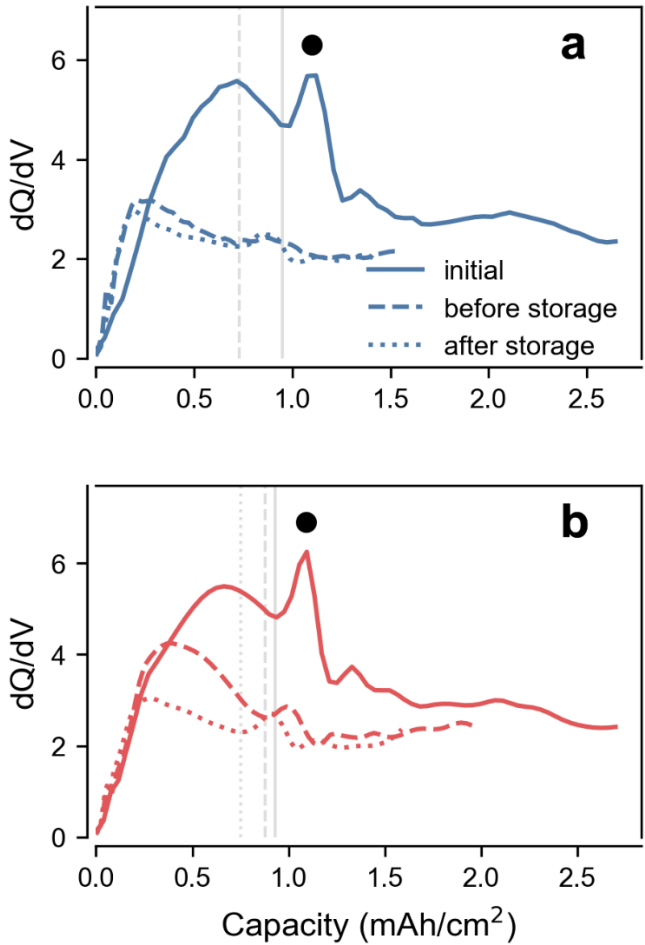

Figure 6. Incremental capacity profiles for *life* (a) and *intermittent* (b) cells at various points of testing. The black circle indicates the peak associated with the onset of graphite lithiation, and vertical lines give visual indication of this onset in each curve. Note that curves are plotted vs. cell capacity, rather than the typical representation vs. voltage. The legend in panel a applies to all panels.

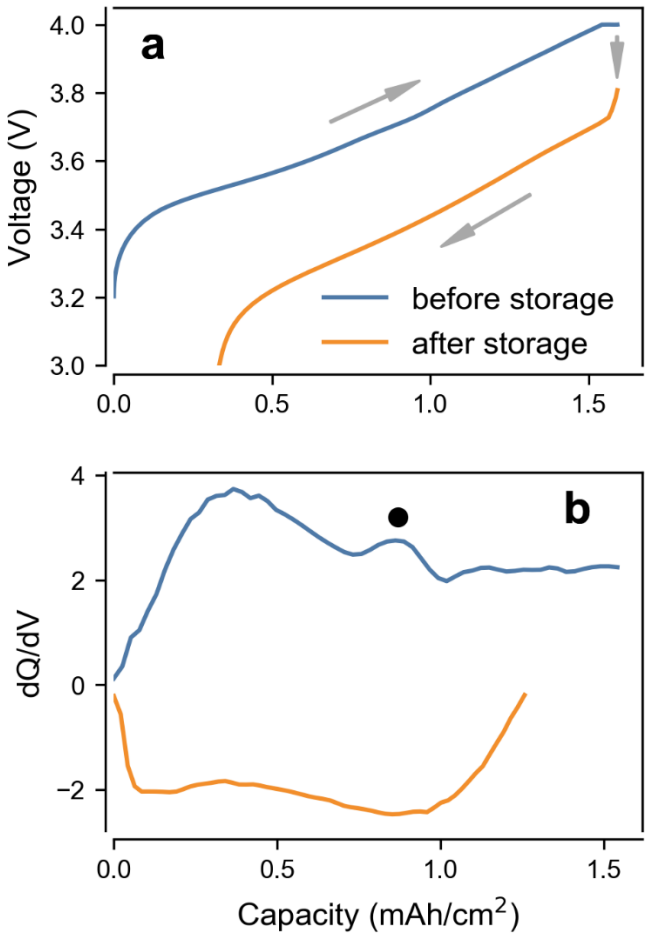

Figure 7. Representative example of the self-discharge behavior of *intermittent* cells after 4 years of storage at 4 V. a) Charge profile immediately before (blue) and discharge profile immediately after (orange) storage. A significant drop in cell voltage is observed during this inactive period. b) Incremental capacity plot (vs. cell capacity) for the two curves shown in panel a. The black circle indicates the peak associated with the onset of graphite lithiation. The color code in panel a applies to all panels.

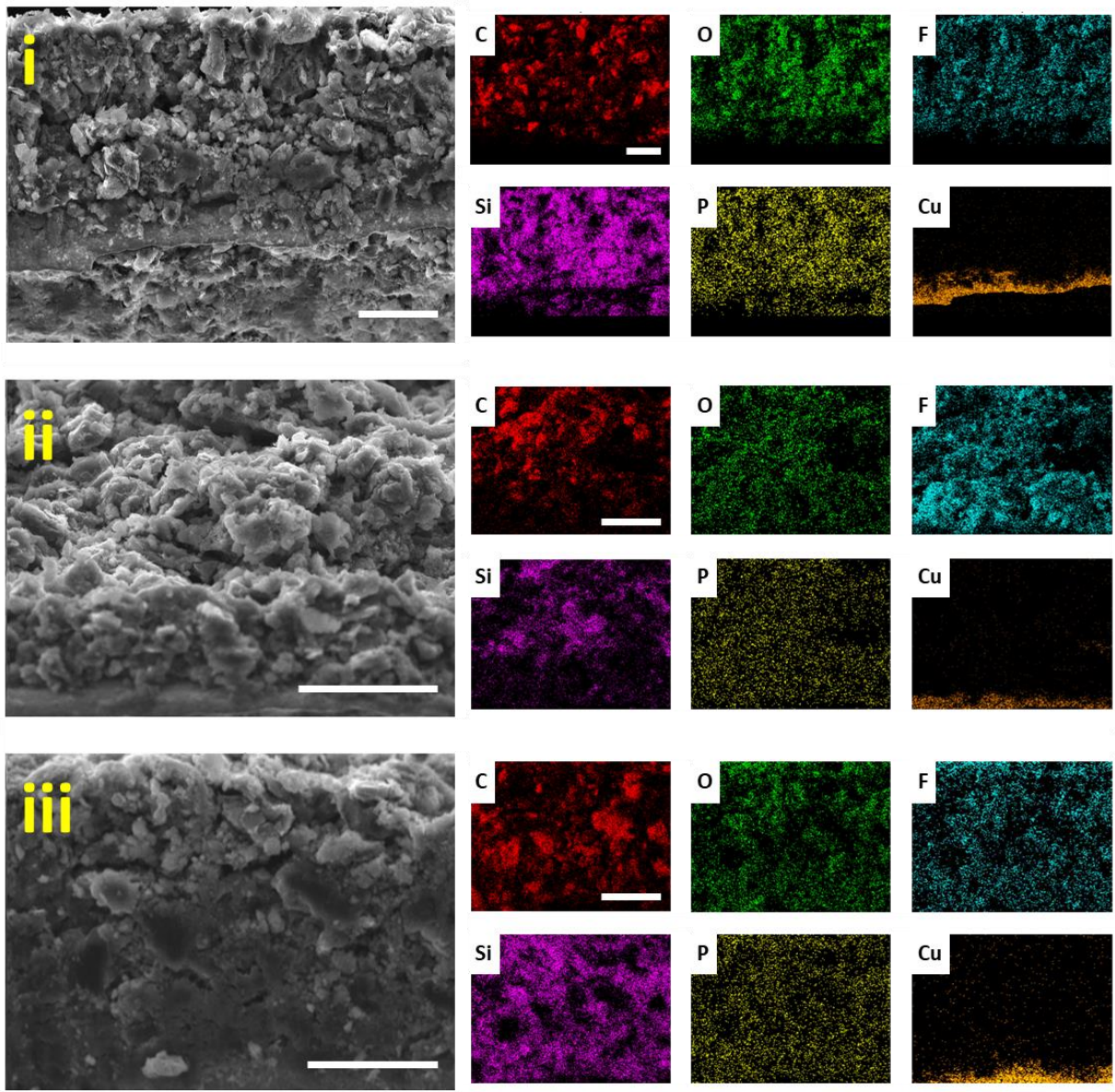

Figure 8. Elemental mapping of the cross-sections of anodes harvested from: i) dry cell. ii) *life* cell. iii) *intermittent* cell. Signal from the active materials (Si and carbon) are detected throughout the entire thickness, suggesting that expansion occurs uniformly throughout the electrode matrix.

# Pouch cells with 15% silicon calendar-aged for 4 years

Marco-Tulio F. Rodrigues[1]*, Zhenzhen Yang[1], Stephen E. Trask[1], Alison R. Dunlop[1], Minkyu Kim[1,2], Fulya Dogan[1], Baris Key[1], Ira Bloom[1], Daniel P. Abraham[1], Andrew N. Jansen[1]

[1] Chemical Sciences and Engineering Division, Argonne National Laboratory, Lemont, IL, USA

2 Current address: Department of Chemistry, Inha University, Incheon, 22212, Republic of Korea

***Corresponding author:** Marco-Tulio F. Rodrigues, marco@anl.gov

***Electrolyte composition.*** Electrolyte samples extracted from *life* and *intermittent* cells were analyzed using HPLC/MS-ESI; before the measurements, samples were treated to remove the LiPF$_6$ salt and volatile species. [R. Sahore et al., Chemistry of Materials, 31 (8), 2884-2891 (2019)] Total-ion chromatograms, and proposed structures for the detected species, are presented in Figure S3. Most products were present in the electrolyte from both types of cells. The only exception was the compounds with m/z = 189, which was only detected in the *intermittent* sample. This signal is tentatively assigned to a highly hydrolyzed fluorophosphate (Figure S3), and its observation in *intermittent* cell agrees with the trends discussed based on XPS results. Although the identity of secondary electrolyte products did not vary much with aging history, differences could be seen in their relative amount. Compounds with high m/z (higher retention time, Figure S3) were slightly more prominent in the *intermittent* sample.

Minor differences between the electrolyte recovered from calendar aged cells were also observed in our NMR studies (data not shown here). The most prominent signals that are not present in the pristine electrolyte were related to alkyl and aldehyde groups. The former is likely related to the many oligomeric species derived from the carbonate solvents, while the latter originates from reactions involving the FEC additive. [Jin et al., Journal of the American Chemical Society, 139 (42), 14992-15004 (2017)] The electrolyte extracted from the dry cell was also analyzed for reference and did not present peaks for the compounds discussed above, indicating that these species continued to form over time. Overall, the compositional similarity between the electrolyte from the aged cells is in agreement with information obtained from XPS; while decomposition reactions progressed with time, they did not show strong dependency on the SOC of storage.

*Supporting Figures*

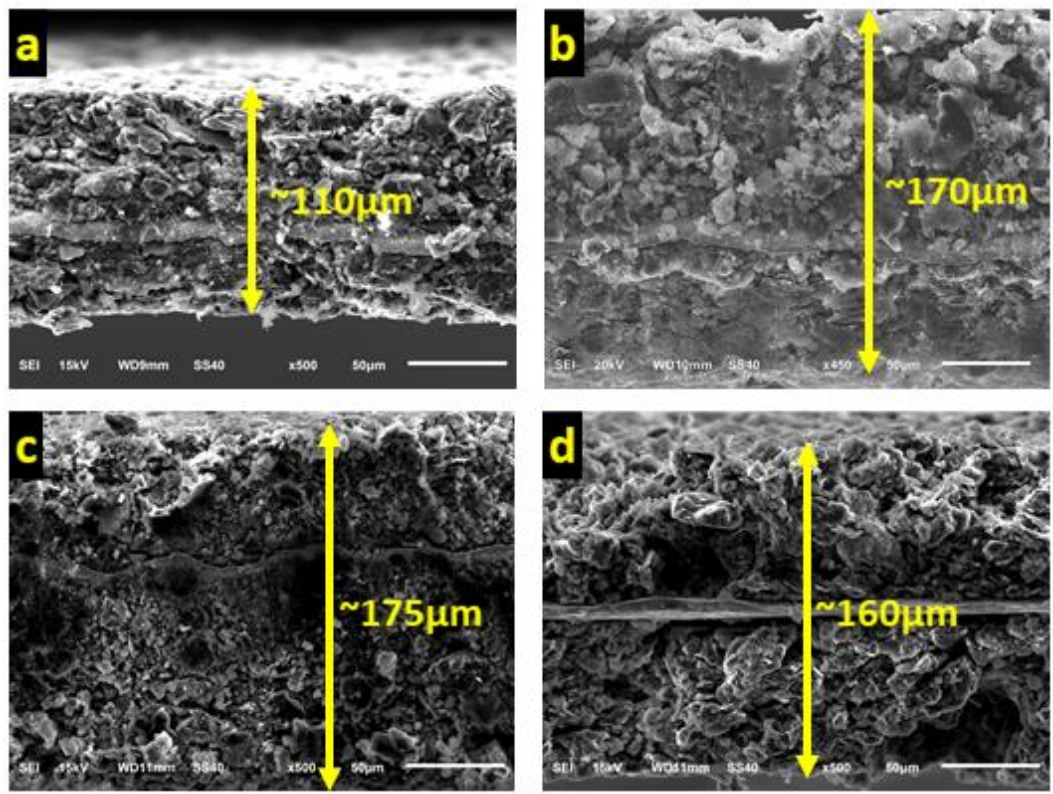

Figure S1. Scanning electron micrographs obtained from the cross-sections of Si-graphite electrodes harvested from the indicated cells. a) uncycled electrode. b) dry cell. c) *life* cell. d) *intermittent* cell. The estimated total thickness of each double-sided electrode is included in the panels. Cycling leads to more irreversible dilation than does storage at high states of charge.

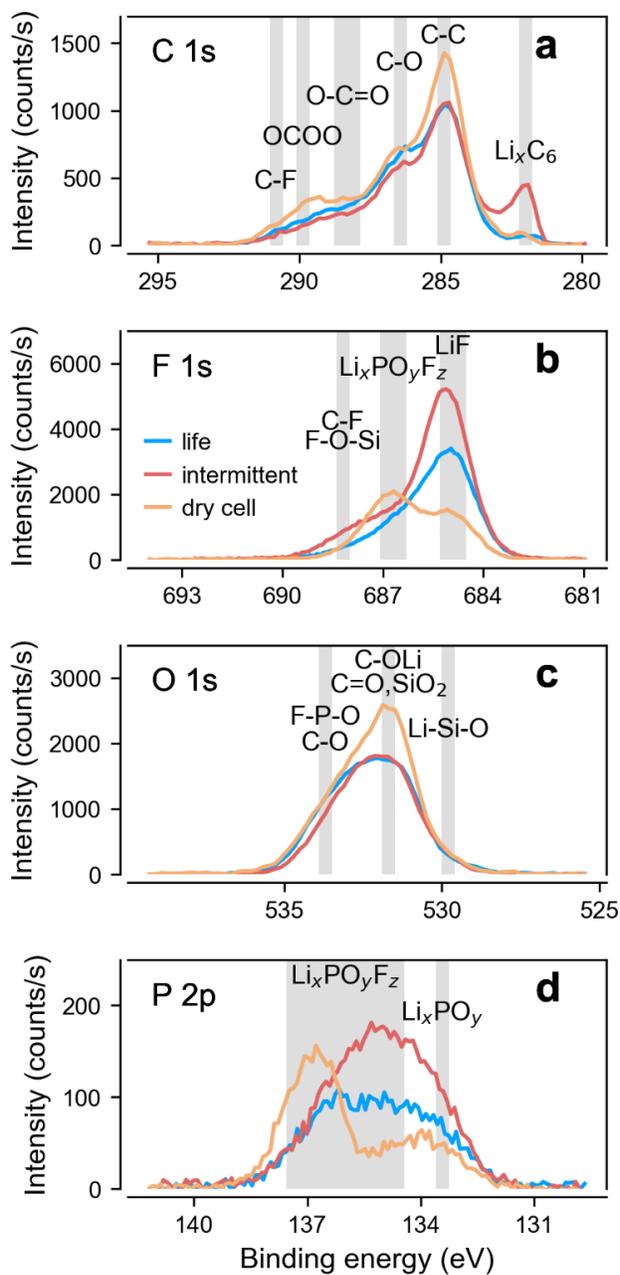

Figure S2. XPS spectra collected from anodes harvested from the dry, *life* and *intermittent* cells. a) Carbon. b) Fluorine. c) Oxygen. d) Phosphorus. The shaded areas indicate the typical binding energies associated with each assigned chemical environment. The color code in panel b applies to all panels.

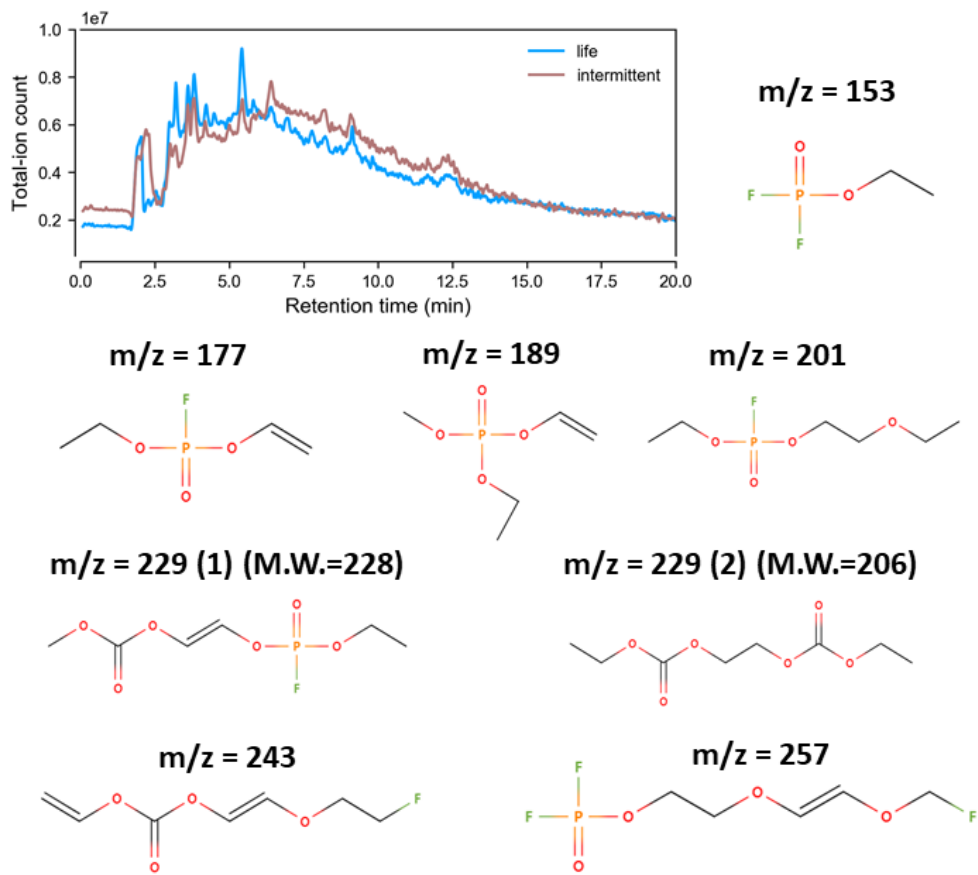

Figure S3. Analysis of electrolyte extracted from *life* and *intermittent* cells. Total-ion chromatograms are shown in the top left, along with tentative structures for all detected compounds. The m/z ratios detected by the mass spectrometer are indicated above each molecular structure.